\documentclass[final,1p,times]{elsarticle} 
\usepackage{graphicx} 
\usepackage{amssymb} 
\usepackage{amsthm} 
\usepackage{lineno} 

\journal{Nuclear Physics A} 
\begin{document} 

\begin{frontmatter} 


\title{J/$\psi$ production in \mbox{p-A} collisions at 158 and 400 GeV: recent results 
from the NA60 experiment}

\author{Enrico Scomparin$^{a}$ for the NA60 collaboration}

\address[a]{INFN Torino, 
Via P. Giuria 1,
Torino, I-10125, Italy}

\begin{abstract} 
The NA60 experiment has studied muon pair production in \mbox{p-A} 
and \mbox{In-In} collisions at the CERN SPS. We present new results on nuclear effects on
J/$\psi$ production in \mbox{p-A} at 158 GeV, the same energy used for collecting \mbox{A-A} data
at the SPS. We then compare nuclear effects with previous results from fixed target experiments, 
and with the results obtained from a \mbox{p-A} data sample taken by NA60 at 400 GeV.
Based on the 158 GeV results, we calculate the expected J/$\psi$ suppression in \mbox{In-In} and
\mbox{Pb-Pb} collisions due to cold nuclear matter effects, and we extract a new estimate for the anomalous
suppression at SPS energies. 
Finally, we show preliminary results on J/$\psi$ polarization in \mbox{p-A} collisions. 

\end{abstract} 

\end{frontmatter} 



\section{Introduction and \mbox{p-A} data analysis}

The suppression of charmonium states in nuclear collisions is considered as one of the most powerful
signatures of the production of a deconfined state~\cite{Sat86}. However, it was soon realized that cold 
nuclear matter
effects, and in particular the interaction of the projectile and target nucleons with the 
$c\overline {c}$ pair, may sizeably contribute to the observed suppression~\cite{Ger88}. 
Such effects are usually studied
in \mbox{p-A} collisions, then extrapolated to \mbox{A-A} and compared with the observed yield in
nuclear collisions, as a function of centrality.

At the SPS, the NA50 experiment has performed an accurate measurement of J/$\psi$ production in \mbox{p-A}
collisions at 400/450 GeV~\cite{Ale06}, i.e. with an incident proton energy higher than the energy per nucleon of
\mbox{Pb-Pb}~\cite{Ale05} and \mbox{In-In}~\cite{Arn07} collisions, studied by the NA50 and NA60 experiments, respectively.
Cold nuclear matter effects have been parameterized by fitting the A-dependence of 
the J/$\psi$ production cross section per \mbox{N-N} collision with the usual $A^{\alpha}$ power-law, or
calculating, in the frame of the Glauber model, the J/$\psi$ absorption cross section 
$\sigma_{abs}^{\rm J/\psi}$. An extrapolation to \mbox{A-A}, based on the assumption of a constant  
$\sigma_{abs}^{\rm J/\psi}$ as a function of incident energy and c.m. rapidity has revealed, by comparison
with J/$\psi$ production yields in \mbox{A-A}, the presence of a suppression which exceeds cold nuclear
matter effects (anomalous suppression). 

In order to provide reference \mbox{p-A} data collected at the same energy and kinematic domain of the
\mbox{A-A} data, NA60 has studied for the first time J/$\psi$ production in \mbox{p-A} collisions at 158
GeV. The incident beam, with an intensity of $\sim 5\cdot 10^8$ protons/s, was sent towards a target 
system made
of several subtargets, with mass numbers ranging from 9 (Be) to 238 (U), which were simultaneously exposed 
to the beam. For details on the NA60 experiment, based on a muon spectrometer coupled to a Si pixel vertex
telescope, see e.g.~\cite{Usa05}.


The analysis of the J/$\psi$ production data at 158 GeV has been performed in the rapidity domain 
$0.28<y_{cm}<0.78$, 
covered with reasonable acceptance by all the sub-targets. The preliminary results shown in this paper
refer to cross-section ratios $\sigma_{pA}^{\rm J/\psi}/\sigma_{pBe}^{\rm J/\psi}$ between the target with 
mass
number $A$ and the lightest one (Be). In this way, the beam luminosity factors cancel out, apart from a
small beam attenuation factor. On the other hand, the track reconstruction efficiencies in the vertex
spectrometer do not cancel out completely, since each target sees the vertex spectrometer under a slightly 
different angle. Therefore, the efficiency of the vertex spectrometer has been computed with the highest
possible granularity (down to the single-pixel level, when track statistics is large enough) and on a 
run-per-run basis. As a check, we have verified, injecting in our 
Monte-Carlo these efficiencies, that we were able to reproduce the muon pair matching rate (of the order 
of $\sim60$\%), and its time evolution, observed for J/$\psi$ events in real data.

\section{Results on \mbox{p-A} collisions}

\begin{figure}[ht]
\centering
\includegraphics[scale=0.33]{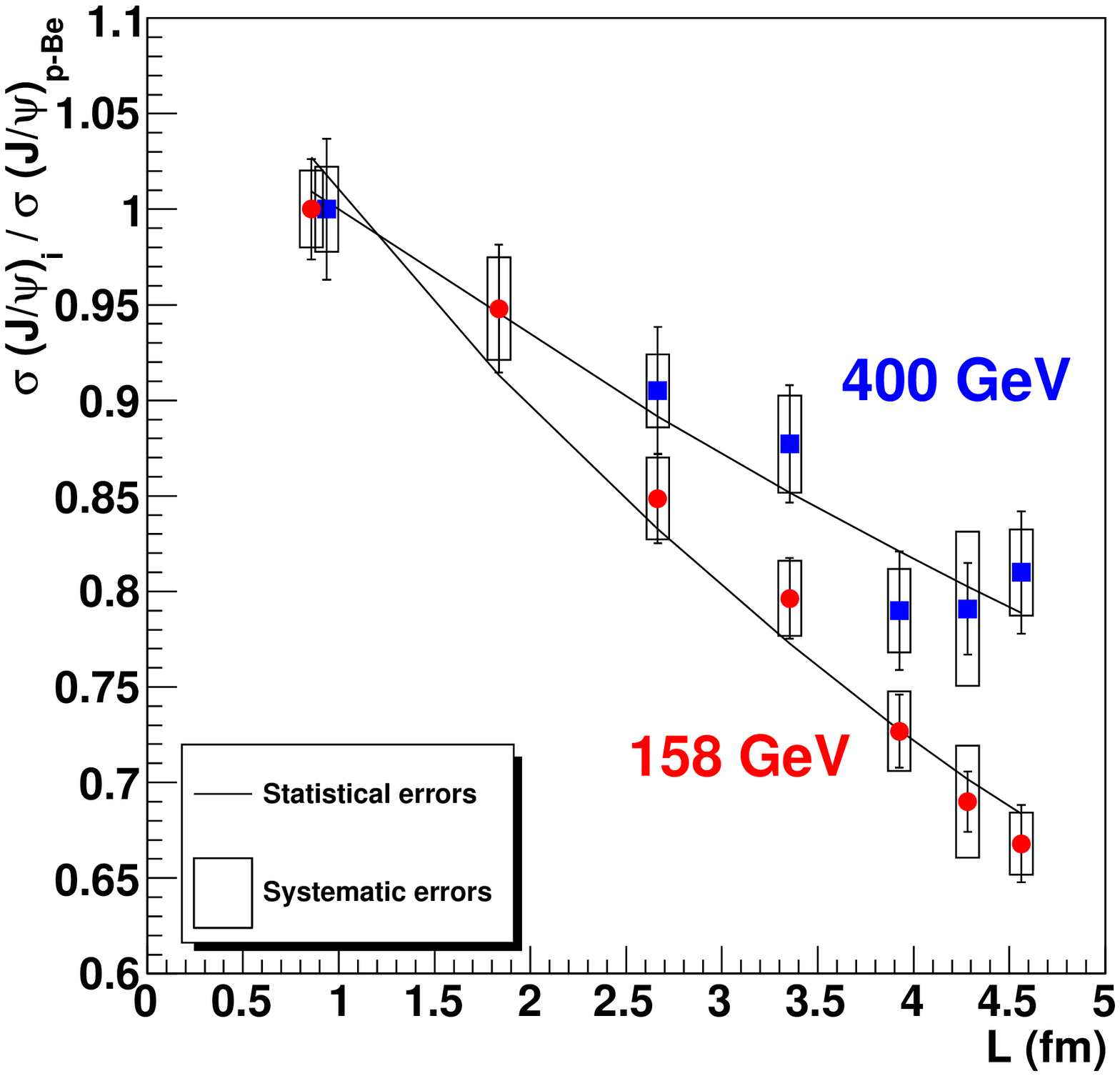}	       
\includegraphics[scale=0.33]{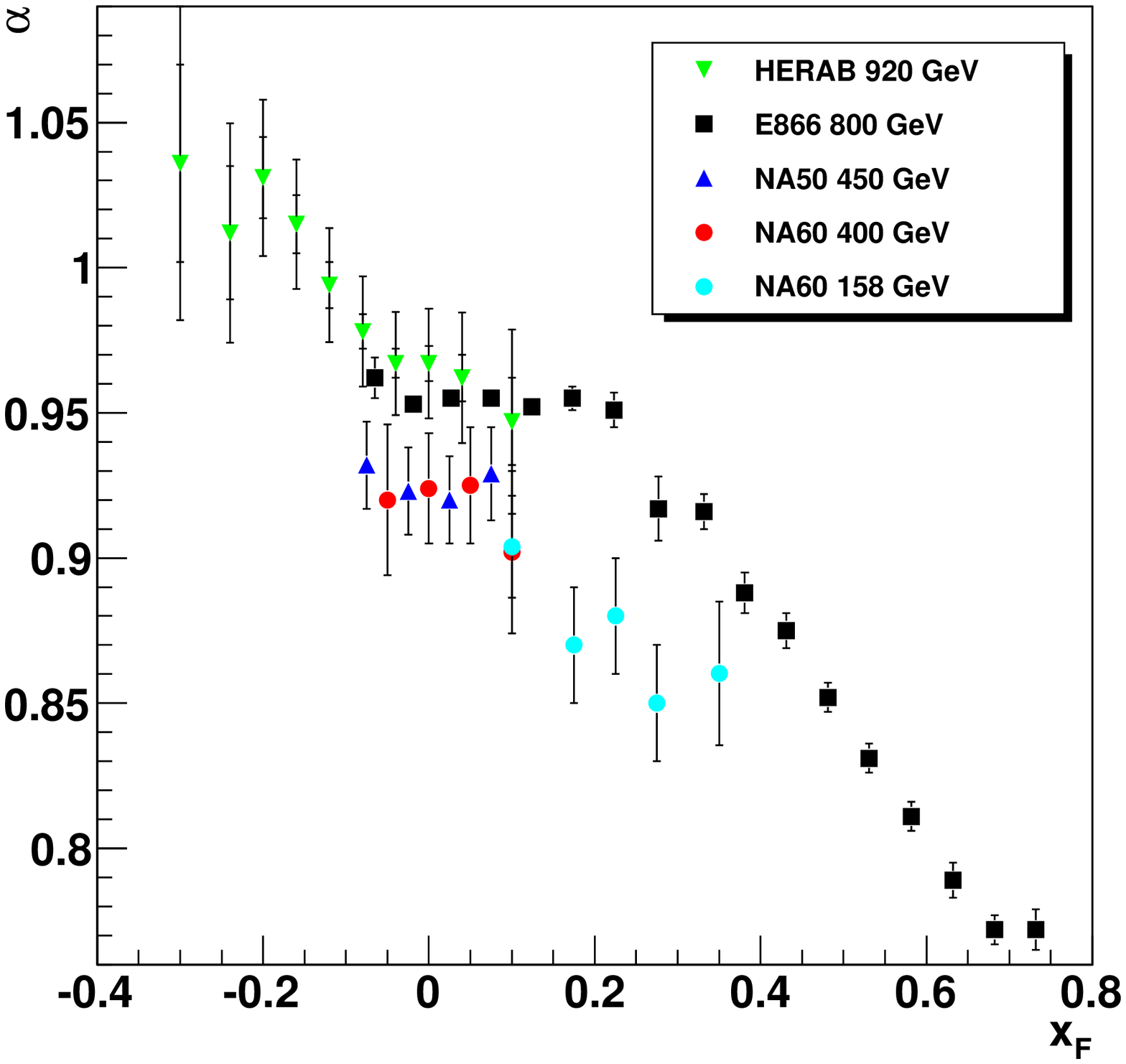}	       
\caption[]{Left: J/$\psi$ cross-section ratios for \mbox{p-A} collisions at 158 GeV (circles) and 400 GeV
(squares), as a function of $L$. Right: compilation of $\alpha$ vs $x_F$.}
\label{fig:1}
\end{figure}

In Fig.~\ref{fig:1}(left) we present the J/$\psi$ cross-section ratios, relative to Be, for the 7 nuclear targets
(Be, Al, Cu, In, W, Pb and U) exposed to the beam. The results are shown as a function of $L$, the mean
thickness of nuclear matter crossed by the $c\overline{c}$ pair in its way through the nucleus. 

The systematic errors shown in Fig.~\ref{fig:1}(left) include contributions from uncertainties on target 
thicknesses, on the $y$ distribution used in the acceptance calculation, and on the reconstruction 
efficiency. We only quote the fraction of the total systematic error which is not common to all the points
(i.e. the one which affects the evaluation of nuclear effects).
By fitting the A-dependence of the cross-section ratios in the frame of the Glauber model, we get 
$\sigma_{abs}^{\rm J/\psi}$(158 GeV)= 7.6$\pm$0.7(stat.)$\pm$0.6(syst.) mb. Alternatively, a fit using the 
$A^{\alpha}$ power-law gives $\alpha$(158 GeV)=$0.882\pm 0.009\pm 0.008$.
In Fig.~\ref{fig:1}(left) we also show the results of the same analysis, carried out on a data sample taken by
NA60 at 400 GeV, with the same configuration of the experimental set-up, in order to minimize the relative
systematic errors. These results refer to the rapidity range $-0.17<y_{cm}<0.33$, corresponding to the same
rapidity in the lab of the 158 GeV data. We clearly note that the A-dependence of this data sample is 
less steep than the one measured at 158 GeV. We get  $\sigma_{abs}^{\rm J/\psi}$(400 GeV)= 
4.3$\pm$0.8(stat.)$\pm$0.6(syst.) mb, and $\alpha$(400 GeV)=$0.927\pm 0.013\pm 0.009$. Nuclear effects on
J/$\psi$ production at 400 GeV had already been studied by NA50, in the range $-0.425 < y_{cm} < 0.575$,
close to the one of the NA60 data. Their result~\cite{Ale06}, 
$\sigma_{abs}^{\rm J/\psi}$(400 GeV)=4.6$\pm$0.6 mb, is in
excellent agreement with our findings.

The observation of a dependence of nuclear effects on the incident proton energy can be further
investigated by comparing our results with previous studies done at fixed target energies. To do that, in
Fig.~\ref{fig:1}(right) we show a compilation of $\alpha$ values as a function of $x_F$, including results
from HERA-B at 920 GeV~\cite{Abt09}, from E866 at 800 GeV~\cite{Lei00} and from NA50 at 450
GeV~\cite{Ale04}. Our analysis at 400 and 158 GeV has
been plotted in Fig.~\ref{fig:1}(right) for various $x_F$ bins. We notice that
nuclear effects become stronger (smaller $\alpha$) at higher $x_{\rm F}$, and that, for a certain 
$x_{\rm F}$, they are also stronger for a lower incident proton energy.
It is worthwhile to note that a theoretical description of the kinematic dependence of cold nuclear matter 
effects on J/$\psi$ production is still missing. An interplay of final state dissociation of the
$c\overline{c}$ pair, parton shadowing and possibly initial state energy loss has been shown to reproduce 
some of the observed features (see e.g.~\cite{Vog00}), but clearly more work is needed in order to arrive 
at a quantitative description.

\section{Anomalous J/$\psi$ suppression in \mbox{In-In} and \mbox{Pb-Pb} collisions}

\begin{figure}[ht]
\centering
\includegraphics[scale=0.33]{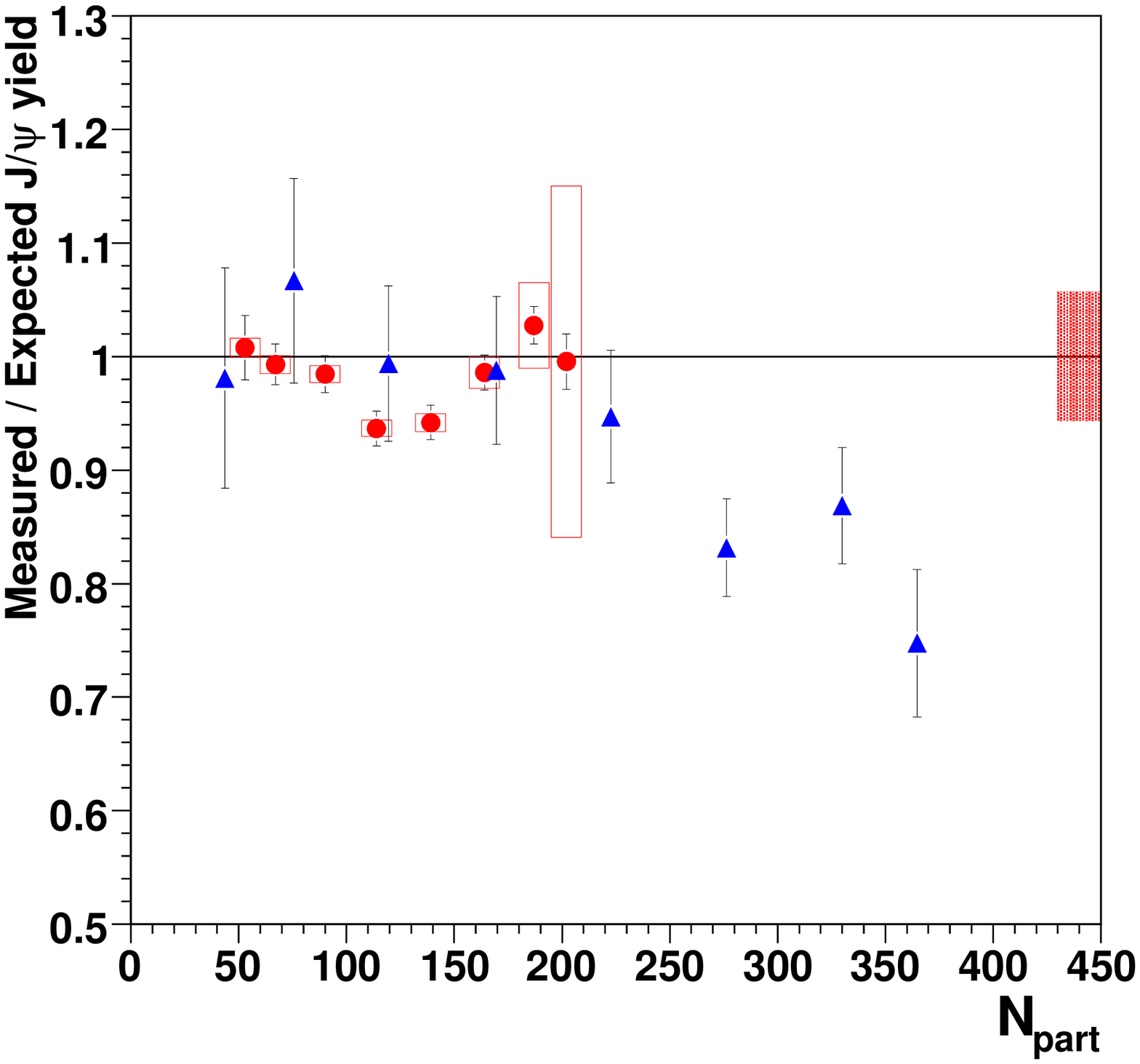}	       
\includegraphics[scale=0.33]{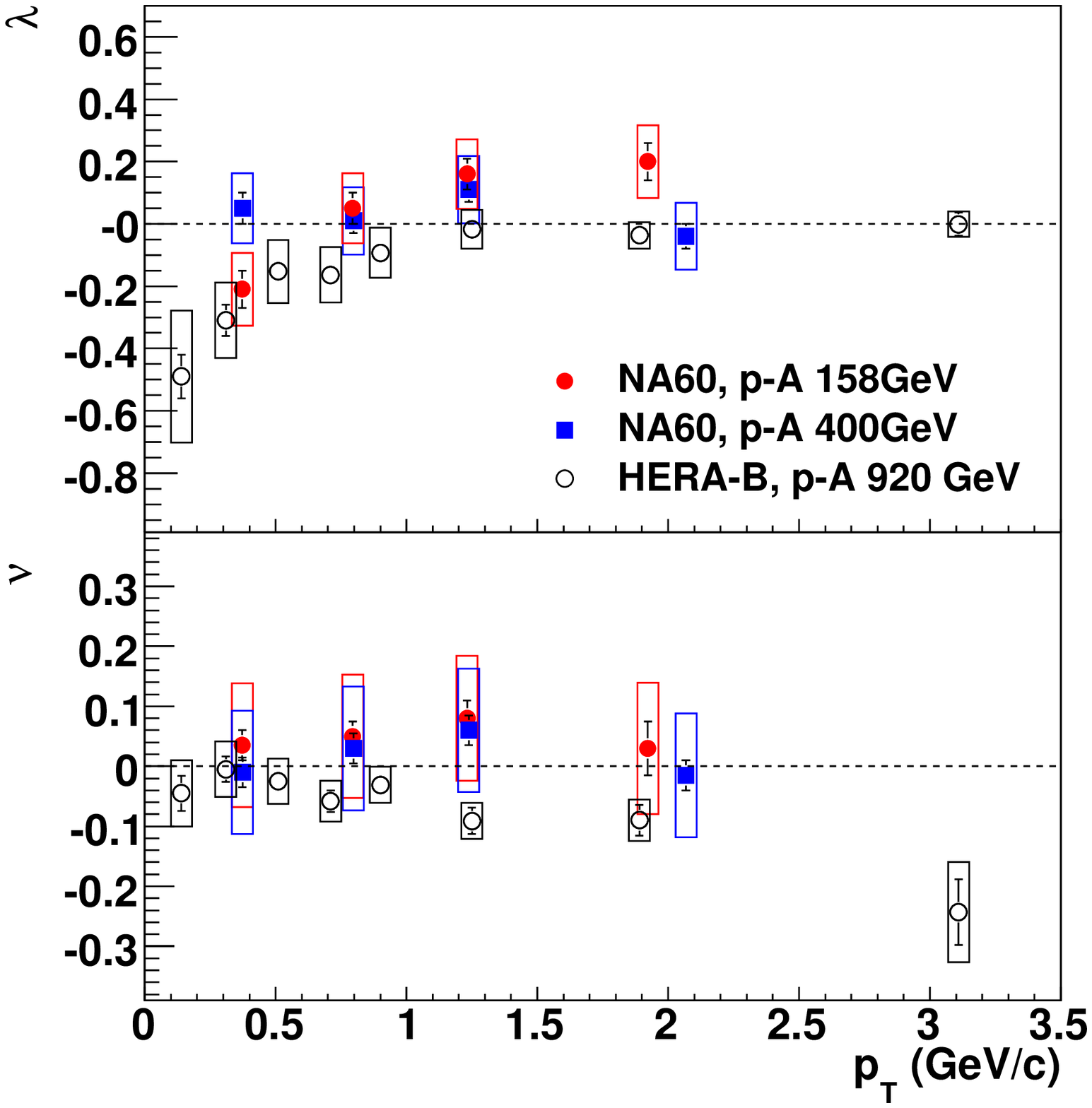}	       
\caption[]{Left: anomalous J/$\psi$ suppression in \mbox{In-In} (circles) and \mbox{Pb-Pb} collisions
(triangles), as a function of $N_{\rm part}$. The boxes around the \mbox{In-In} points represent 
correlated systematic errors. The filled box 
corresponds to the uncertainty in the absolute normalization of the \mbox{In-In} points. A 12\% global
error, due to the uncertainty on $\sigma_{abs}^{\rm J/\psi}$(158 GeV), is not shown.
Right: the J/$\psi$ polarization parameters $\lambda$ and $\nu$, in the helicity reference frame, as a function of 
$p_{\rm T}$, for NA60 data, compared with recent results from HERA-B. The boxes represent the total 
errors.}
\label{fig:2}
\end{figure}

The \mbox{p-A} results at 158 GeV shown in the previous section have been collected at the same energy and
in the same $x_{\rm F}$ range of the SPS \mbox{A-A} data.  
It is therefore natural to use these results to calculate the expected size of cold nuclear matter effects
on J/$\psi$ production in nuclear collisions. In order to do that, we have determined, as a function of
the forward energy $E_{\rm ZDC}$ and using the Glauber model, the expected shape 
$dN^{expec}_{{\rm J}/\psi}/dE_{\rm ZDC}$, assuming that J/$\psi$ production scales with the number of 
\mbox{N-N} collisions and that the produced J/$\psi$ are absorbed in nuclear matter according to 
the value $\sigma_{abs}^{\rm J/\psi}$(158 GeV) given in the previous section. 
The measured $dN_{{\rm J}/\psi}/dE_{\rm ZDC}$ has then been 
normalized to $dN^{expec}_{{\rm J}/\psi}/dE_{\rm ZDC}$ using the procedure detailed in Ref.~\cite{Arn07}.
This procedure, which was used up to now, does not take explicitly into account, when extrapolating from 
\mbox{p-A} to \mbox{A-A}, the presence of shadowing effects. It can be shown~\cite{Arn09} that in the 
kinematic region 
where \mbox{A-A} data are measured ($0<y_{cm}<1$), this method leads to an overestimation of the anomalous
suppression, of the order of $\sim 5\%$. We have therefore corrected our result for this small bias, using
the EKS98~\cite{Esk99} parameterization of shadowing effects.
In Fig.~\ref{fig:2}(left) we present, as a function of the number of participants, our result for the 
anomalous J/$\psi$ suppression in \mbox{In-In} and \mbox{Pb-Pb} collisions.
We can see that up to $N_{part}\sim 200$ the J/$\psi$ yield is, within errors, compatible with our
extrapolation of cold nuclear matter effects. For $N_{part}> 200$ an anomalous suppression is present, which
reaches $\sim20-30\%$ for central \mbox{Pb-Pb} collisions. With this new evaluation of the anomalous
suppression, the effect becomes smaller with respect to the past. This is essentially due to the
larger $\sigma_{abs}^{\rm J/\psi}$ value now used in the determination of the nuclear reference.


\section{J/$\psi$ polarization in \mbox{p-A} collisions}

A study of the J/$\psi$ polarization in \mbox{p-A} collisions can be performed by studying the angular
distribution of the decay muons. This study has been shown to be relevant, at collider energies, for 
investigating the quarkonium production models, since various theoretical approaches~\cite{Lan09} predict 
different values, as a function of $p_{\rm T}$, for the polarization parameters $\lambda$, $\mu$, $\nu$, 
obtained through a fit of the muon angular distribution $d^2\sigma/d\cos\theta d\phi \propto 
1+\lambda\cos^2\theta+\mu\sin 2\theta\cos\phi+(\nu/2)\sin^2\theta\cos 2\phi$. In
Fig.~\ref{fig:2}(right) we present our preliminary results for $\lambda$ and $\nu$ ($\mu$ is compatible
with zero everywhere), obtained in the helicity
reference frame, compared with recent HERA-B
results~\cite{Abt092}. The data seem to indicate slightly negative $\lambda$ values at low $p_{\rm T}$,
which level around zero at larger transverse momentum. $\nu$ values are close to zero in the 
$p_{\rm T}$ range explored by NA60.

\section {Conclusions}

We have shown new results on J/$\psi$ production in \mbox{p-A} collisions at 158 and 400 GeV. We
see that nuclear effects become more important when moving towards lower energy, an observation that
remains valid when extending the comparisons to other sets of results. Using the new 158 GeV results for
determining the expected cold nuclear matter effects in \mbox{A-A} results in a smaller anomalous 
suppression with respect to previous estimates. The effect is anyway still sizeable ($\sim$25\%) for central
\mbox{Pb-Pb} collisions.




\begin{thebibliography}{00} 
   
\bibitem{Sat86} T.~Matsui and H.~Satz, {\it Phys. Lett.} {\bf B178} (1986) 416.
\bibitem{Ger88} C.~Gerschel and J.~Huefner, {\it Phys. Lett.} {\bf B207} (1988) 194.
\bibitem{Ale06} B.~Alessandro et al. [NA50 Collaboration], {\it Eur. Phys. J.} {\bf C48} (2006) 329. 
\bibitem{Ale05} B.~Alessandro et al. [NA50 Collaboration], {\it Eur. Phys. J.} {\bf C39} (2005) 335.
\bibitem{Arn07} R.~Arnaldi et al. [NA60 Collaboration], {\it Phys. Rev. Lett.} {\bf 99} (2007) 132302.
\bibitem{Usa05} G.~Usai et al. [NA60 Collaboration], {\it Eur. Phys. J.} {\bf C43} (2005) 415.
\bibitem{Abt09} I.~Abt et al. [HERA-B Collaboration], {\it Eur. Phys. J.} {\bf C60} (2009) 525.
\bibitem{Lei00} M.J.~Leitch et al. [E866 Collaboration], {\it Phys. Rev. Lett.} {\bf 84} (2000) 3256.
\bibitem{Ale04} B.~Alessandro et al. [NA50 Collaboration], {\it Eur. Phys. J.} {\bf C33} (2004) 31.
\bibitem{Vog00} R.~Vogt, {\it Phys. Rev.} {\bf C61} (2000) 035203. 
\bibitem{Arn09} R.~Arnaldi, P.~Cortese and E.~Scomparin, [arXiv:0909.2199].
\bibitem{Esk99} K.J.~Eskola, V.J.~Kolhinen and C.A.~Salgado, {\it Eur. Phys. J.} {\bf C9} (1999) 61.
\bibitem{Lan09} J.P.~Lansberg, [arXiv:0811.4005v1].
\bibitem{Abt092} I.~Abt et al. [HERA-B Collaboration], {\it Eur. Phys. J.} {\bf C60} (2009) 517.


\end{thebibliography}
\end{document}